\documentstyle[aps,epsf]{revtex}
\def\lesssim{\mathrel{\raise.3ex\hbox{$<$\kern-.75em\lower1ex\hbox{$\sim$}}}}
\begin{document}
\title{Non-Equilibrium Steady States and Transport\\
  in the Classical Lattice $\phi ^4$ Theory} 
\author{Kenichiro Aoki\footnote{E--mail: {\tt ken@phys-h.keio.ac.jp}}
  and Dimitri Kusnezov\footnote{E--mail: {\tt
      dimitri@nst.physics.yale.edu}}}
  \address{$^*$Dept. of
  Physics, Keio University, {\it 4---1---1} Hiyoshi, Kouhoku--ku,
  Yokohama 223--8521, Japan\\ 
  $^\dagger$Center for Theoretical Physics,
  Sloane Physics Lab, Yale University, New Haven, CT\ 06520-8120}
\date{\today } \maketitle

\begin{abstract}
  We study the classical non-equilibrium statistical mechanics of
  scalar field theory on the lattice. Steady states are analyzed near
  and far from equilibrium.  The bulk thermal conductivity is
  computed, including its temperature dependence.  We examine the
  validity of linear response predictions, as well as properties of
  the non-equilibrium steady state.  We find that the linear response
  theory applies to visibly curved temperature profiles as long as the
  thermal gradients are not too strong.  We also examine the
  transition from local equilibrium to local non-equilibrium.
\end{abstract}
\newpage

The understanding of the dynamics of non--equilibrium field theories
is important to many areas in physical sciences,  from
processes in inflation or baryogenesis in the early universe, 
transport processes in condensed matter, to the  possible
states of hadronic matter in heavy ion collisions, such as
quark--gluon plasma, disoriented chiral condensates and color
superconducting states.  
However, many problems linger even in the basic understanding of
non-equilibrium statistical mechanics and transport\cite{revs}.
It is clearly desirable to address these problems regarding the
non-equilibrium dynamics of field theories while making as few
assumptions about the dynamics of the theory as possible.  In
this work, we study the steady state dynamics of classical
massless $\phi^4$ lattice field theory in $(d+1)$ dimensions
($d=1,3$), under weak and strong thermal gradients.  We study
the classical field theory on a lattice since the problems are
well defined and techniques are available to construct
non-equilibrium states from first principles.  It is not clear
to us how to address the questions we pose in the full
quantum theory without making some drastic assumptions.  
Our computations are non--perturbative.
Furthermore, not only is the temperature within the
system dynamical, but even the question of whether local
equilibrium is achieved is {\it not} an assumption but is
determined {\it dynamically} by the system.  The equilibrium
properties of classical $\phi^4$ theory have been studied in the
past, including its ergodic properties\cite{gong,parisi},
Hamiltonian dynamics and phase transitions\cite{ssb}. However,
the kind of questions we address here have not been
answered in the previous literature.

Near equilibrium, linear response theory supposedly holds; yet
there is no means to address its regime of validity within the
theory itself since the computation is performed in equilibrium.
By explicitly constructing non-equilibrium steady states near
equilibrium,
we examine the validity of the linear response theory.  We
then construct steady states of the system under stronger
thermal gradients and study their physical properties. Here,
fundamental questions arise, such as under which conditions
local equilibrium is achieved.  Local equilibrium, an assumption
that equilibrium concepts can be applied locally to a problem
which might be globally non-equilibrium, is widely
used\cite{mclennan}. In fact, thermalization is a concept
which is assumed, often tacitly, in many applications of
non-equilibrium physics.  However, as a system moves away from
equilibrium, what precisely constitutes `local equilibrium'
becomes unclear and it is of interest to understand what kind of
deviations develop from it and why.  These problems are also
non-trivial from the statistical mechanical point of view; even
small departures from equilibrium into non-equilibrium steady
states (such as the those we study) already lead to peculiar
behaviors including a divergent Gibbs entropy $S_G$
($S_G\rightarrow - \infty$) and a multi-fractal steady state
measure\cite{non-eq,revs}.  The steady-state distributions far
from equilibrium are not well understood classically and very
little is known concerning their quantum counterparts.  We would
like to see to what extent these peculiar statistical measures
influence the steady state thermal profiles, $T(x)$.

Classical field theories are relevant to the high temperature
dynamics of quantum field theories and have proved effective,
for instance, in computing finite temperature properties of the
standard model \cite{sm,gong,classical-ym}.  Furthermore,
non-equilibrium dynamics of classical field theories is of
interest in its own right, an understanding of which is
essential to understanding the dynamics of the quantum theory.
The approach we adopt here can be applied to classical lattice
theories quite generally.  Transport properties have been
previously studied in scalar quantum field theory using linear
response theory\cite{hosoya}\ and the theory is known to have a
classical, finite temperature limit for correlation
functions\cite{smit}.  Yet the question of how to relate our
results to those results is far from trivial and will not be
pursued here.

We start with the Lagrangian
\begin{equation}  \label{continuum-lagrangian}
  -{\cal L}=
  {\frac{1}{2}} \left(\frac{\partial\tilde\phi (\tilde x)} 
   {\partial\tilde x_\mu}
  \right)^2 +{\frac{\tilde g^2}{4}} \tilde\phi(\tilde x)^4 .
\end{equation}
This model, when discretized, reduces
to a model of lattice vibrations with quartic anharmonicity,
with the following dimensionless Hamiltonian
\begin{equation}  \label{lattice-hamiltonian}
  H(\pi,\phi)=\frac{1}{2}\sum_{\bf r}\left[\pi_{\bf r}^2 +
    \left(\nabla \phi_{\bf r}\right)^2 + \frac{1}{2}\phi_{\bf r}^4\right].
\end{equation}
Here $\pi=\partial \phi/\partial t$, ${\bf r}$ runs over all
sites in the lattice, and the lattice derivative has components
$\nabla_k\phi_{\bf r}\equiv \phi_{\bf r + e_k}-\phi_{\bf r}$
(${\bf e_k}$ is the unit lattice vector in the $k$-th
direction). The two theories are related by discretization and
the rescalings, $\phi_{\bf r}(t)=a\tilde g\tilde \phi(\tilde
{\bf r},\tilde t ), \quad t=\tilde t/a,\quad {\bf r}=\tilde {\bf
  r}/a $, where $a$ is the lattice spacing. The equations of
motion, $\Box\phi = -\phi^3$, are solved on a spatial grid,
using two methods: fifth and sixth order Runge-Kutta, and
leap-frog algorithms\cite{numrec}.

In order to generate a stationary non-equilibrium statistical
ensemble, thermal boundary conditions are imposed on the
equations of motion.  Specifically, at $x=0$ and $x=L$, we add
two time-reversal invariant fields which act to dynamically
thermalize these boundaries at given temperatures $T_1$ and
$T_2$ \cite{thermostats} (a more detailed account will be given
elsewhere).  For $d>1$ we impose periodic boundary conditions on
the other directions.  Apart from the thermal boundary
conditions, the system evolves according to the dynamics
dictated by the Hamiltonian (\ref{lattice-hamiltonian}). We used
from $10^6$ to $10^9$ time steps of $dt$ from $0.1$ to $0.001$,
with observables being sampled every $\Delta t=20\sim100\,dt$.
In $d=1$, the lattice size was varied from $L=20$ to 8000, while
in $d=3$ it ranged from $50\times N\times N$ ($N$ ranging from
3--20) to $1000\times 3\times 3$. We have verified that when
$T_1=T_2$, these boundary conditions dynamically set all the
temperatures inside the system to be equal to the boundary
temperatures and reproduce the equilibrium canonical measure
$\rho_{eq}(\pi,\phi)\sim \exp[-H(\pi,\phi)/T_1]$ at all points.

By controlling $T_1$ and $T_2$ we can begin to explore the
non-equilibrium steady state. One question we would like to
address is how the temperature profile $T(x)$ behaves. Near
equilibrium one would expect a linear profile, but beyond that
the shape is unknown. In our near equilibrium simulations,
$T_1\lesssim T_2$, we find a linear temperature profile and
recover transport given by Fourier's law, as shown in
Fig. 1~(a). However, far from thermal equilibrium ($T_1\ll
T_2$), the temperature profile develops significant curvature,
seen in 1~(b),(c) for the ratio $T_2/T_1=10,20$.  It is of
interest to understand the physics behind these temperature
profiles.  We would further like to understand until what point
linear response and local equilibrium provide reasonable
descriptions.

\underline{\sl Equilibrium: $T_1=T_2$.} A standard approach to
thermal conductivity utilizes equilibrium correlation functions
to compute near-equilibrium transport. This linear
response approach uses the Green--Kubo formula,
\begin{equation}  \label{green-kubo}
  \kappa(T)={\frac{1}{T^2}}
  \int_0^\infty\!\!\!\!dt\int\!\!d{\bf r}\,
  \left\langle {\cal T}^{0x}({\bf r},t)
  {\cal T}^{0x}({\bf r}_0,0)\right\rangle_{eq} ,
\end{equation}
where the autocorrelation function is evaluated in the canonical
ensemble, $T_1=T_2$. For our lattice calculation, we replace $\int
d{\bf r}$ with a lattice sum.  It is interesting to note that the
integrand in (3) has been argued to develop a long time tail behavior
of $\sim t^{-d/2}$, leading to the divergence of (3) in $d=1$
\cite{tail}.  In Fig. 2~(top), a typical autocorrelation function for
(1+1) dimensions is plotted to several hundred times the mean free
time, which is well into the regime where long-time tails would be
evident. The time integral is given in Fig.~2~(bottom), showing that the
integral (3) is finite, which we attribute to the `on-site' nature of
the $\phi^4$ interaction, in contrast to some of the other models
\cite{others}.  Interestingly enough, we do find that the 
transient behavior of the Green--Kubo integrand is quite close to
$t^{-1/2}$ up to a few ten times the mean free time, after which it
decays much faster.  Consistent results are found in $d=3$ as well.
The computed $\kappa(T)$ is shown in Fig.~3 (top).

\underline{\sl Near Equilibrium: $T_1\lesssim T_2$.} Near equilibrium,
we find that a linear temperature profile emerges dynamically.  The
thermal conductivity $\kappa$ is then obtained through Fourier's law:
\begin{equation} 
  \label{eq:tc}
  \kappa(T) = -\frac{\langle {\cal T}^{0x}
           \rangle_{{\scriptscriptstyle NE}}} {\nabla T},\qquad
         {\cal T}^{0x}=-\partial_t\phi\nabla\phi,  
\end{equation}
where $\langle {\cal T}^{0x}\rangle_{{ \scriptscriptstyle NE}}$
is the heat flux averaged over the non-equilibrium steady state.
Attention is paid to verifying the linear response properties by
varying the temperature difference $|T_2-T_1|$ around the same
average temperature. $T(x)$ is the local temperature defined
through an ideal gas thermometer, by
$T(x)=\langle\pi^2(x)\rangle_{{ \scriptscriptstyle NE}}$, where
$\pi(x)$ is the momentum density.  This will serve as a
convenient definition as long as local equilibrium is achieved
and the momentum distributions are gaussian.  Here
$\langle\cdots\rangle_ {{ \scriptscriptstyle NE}}$ indicates the
ensemble average over the non-equilibrium steady state. To
obtain the transport properties, each simulation is run long
enough for observables such as ${\cal T}^{0x}$, the energy
density as well as distribution functions to converge.

In Fig. 3~(top), we compile the Green-Kubo and direct measurements of
$\kappa$, plotted as a function of $T$.  We find that these
independent computations are quite consistent with each other.
$\kappa$ is found to have a temperature dependence
\begin{equation}
 \kappa=AT^{-\gamma},\qquad \cases{
   \gamma=1.38(2),\  A=2.72(4) & (1+1) dimensions\cr
   \gamma=1.64(4),\  A=9.1(2)  & (3+1) dimensions \cr}.
\end{equation}
This behavior is similar to that of lattice phonons at high
temperature\cite{phonon}. We have also verified that a sensible
bulk behavior exists, as shown in Fig. 3~(bottom); the thermal
conductivity is independent of $L$ when it is larger than the
mean free path, which, on the lattice, is of order of the
conductivity.

In trying to understand  near equilibrium physics,
one might be tempted to assign a statistical measure, such as
$\rho_{\scriptscriptstyle NE}(\pi,\phi)\sim \exp[-
H(\pi,\phi)/T(x)]$ to the non-equilibrium stationary state, or
similar measures which 
assume some form for $T(x)$. Strictly speaking, this is not
correct; the phase space measures which describe steady state
non-equilibrium systems are multi-fractal (whether one studies
shearing, heat flow and so forth), converging to  smooth
distributions (Boltzmann) only in the equilibrium
limit\cite{non-eq,revs}. An important consequence is that the
dynamical space is necessarily of lesser dimension than the
equilibrium phase space. This in turn results in additional
correlations of $\pi(x)$ and $\phi(x)$. One can also see that the
non-equilibrium measure is not locally Boltzmann since
quantities such as $\langle\pi(x)\phi(x')\rangle\not=0$ for
$x\not=x'$ which is also reflected in the non-zero energy
flow, while $\langle\pi(x)\rangle=0$ near and far from
equilibrium.

\underline{\sl Far From Equilibrium: $T_1\ll T_2$.} When the
temperature gradients are larger and we are no longer in the linear
regime, the temperature profile becomes visibly curved.
An example of such a non--linear temperature profile are given in
Fig.~1.  It should be noted that inside the boundaries, the dynamics
is that of only the $\phi^4$ theory and the temperature profile is
determined by it.  When the temperature varies substantially in the
system, one cause for the non--linearity is the temperature dependence
of the thermal conductivity. If this were the only cause of
non--linearity, the temperature profile can be determined within the
region by integrating Eq. (4) (when $\gamma\neq1$)
\begin{equation} \label{t-profile}
  T(x) = T_1\left[1-\left(1-\left(\frac{T_2}{T_1}
      \right)^{1-\gamma} \right)
    {\frac{x}{L}}\right]^{{\frac{1}{1-\gamma}}}.
\end{equation}
$T(x)$ is a function only of $x/L$ so that it is consistent with a
smooth continuum limit.  This expression for $T(x)$ provides very good
descriptions of the measured profiles with significant
curvature, as can be seen in Fig. 1 where Eq. (6) (dashes) is
almost indistinguishable from the measured steady state profiles
(solid). This
indicates that linear response extends well beyond the regimes of
small temperature differences, when applied locally.  {\it \`A
  priori}, this was not clear.

By continuing to increase the ratio $T_2/T_1$ in the simulation,
we do eventually reach steady state situations where the linear
response formula no longer works. At this point, we begin to see
indications that the concept of local equilibrium also becomes
more tenuous.  To analyze these questions more concretely, we
need a dimensionless measure of how strong the thermal gradient
is.  A natural choice we adopt is $\lambda \nabla T/T$, where
$\lambda$ is the mean free path.  In our model, the heat
capacity per unit volume, $C_V$, and the sound speed, $c_s$, are
of order unity, so that elementary kinetic theory suggests that
the mean free path is $\lambda\sim d\kappa$, where $d$ is the
spatial dimension. To quantify the departures from the linear
response formula, we plot in Fig. 4 (for one spatial dimension)
the deviation of the measured heat flux to that obtained from linear response
prediction using Eqs.~(\ref{eq:tc})---(\ref{t-profile}) (denoted 
$\langle{\cal
T}^{0x}\rangle_{NE}$ and $\langle {\cal T}^{0x}\rangle_{LR}$, respectively),
as a function of the
quantity $\kappa(T) \nabla T/T$.  The figure includes different
lattice sizes and temperatures.  We see that for
\begin{equation}
  \label{eq:condition}
  \kappa(T) \frac{\nabla T}{T}\ll 1
\end{equation}
linear response theory holds quite well. This includes systems with
significant curvature in $T(x)$ such as Fig. 1. Eventually, for
sufficiently strong gradients, the measured heat flow begins to
deviate from the linear response results; the system does not
conduct heat as well as its linear response theory prediction.
At this point, we observe simultaneously the departure of other
quantities from a `local equilibrium' characterized by   gaussian momentum 
distributions.  To see the departure from
local equilibrium, we follow the behavior of various
observables, which include the momentum cumulants, the
steady-state momentum distributions $f(\pi_k)$ as well as heat
flux and correlation functions.  For instance, in equilibrium,
$\langle\pi^4(x)\rangle/\langle \pi^2(x)\rangle^2= 3$,
$\langle\pi^6(x)\rangle/\langle \pi^2(x)\rangle^3= 15$ and so
forth.
In the regime where the linear response theory breaks down, the
momentum distributions become more sharply peaked and are no
longer gaussian even in the steady state, and the deviations in
the cumulants become apparent.
This indicates that at least in our theory, higher order
corrections to linear response are not that well founded since the
concept of temperature becomes tenuous at this point.
Since the heat flow is a constant, we also note that the local equilibrium
condition (\ref{eq:condition}) is more likely to be satisfied at the
higher temperature end.  Similar behavior is seen in (3+1) dimensions.

We have further examined the behavior of the (coarse grained) Boltzmann
entropy $S_B$.  While Gibbs entropy is known to be singular, any similar 
divergence in the coarse grained $S_B$ would only be evident if
one extrapolated the measured values to the continuum limit. 
Instead, we consider if the notion of  local
entropy changes significantly as the system moves away from
equilibrium. To this end, we compute the Boltzmann entropy $S_B$
from the 1- and 2-body densities $f^{(1)}(\pi(x),\phi(x))$ and
$f^{(2)}(\pi(x),\phi(x),\pi(x'),\phi(x'))$ in the
non-equilibrium steady states, from $S^{(k)}_B=-\int d\mu^{(k)}\,
f^{(k)}\log f^{(k)}$.  We find that $S^{(1)}$ does not shift noticeably from
its equilibrium value regardless of how far the system is from
equilibrium. Further, $S^{(2)}$ $(\alt 2 S^{(1)})$ is only
slightly less than its upper limit $2 S^{(1)}$ and remains so
even far from equilibrium. So unlike $S_G$ $\left(\leq V
  S^{(1)}_B\right)$, $S_B$ is found to be rather insensitive to
the non-equilibrium nature of the system. While this could be a
manifestation of coarse graining, it does suggest that
some local thermodynamic concepts might still be useful in making
connections with non-equilibrium thermodynamics.

The behavior of scalar lattice field theory near and far from
equilibrium has been explored, and the present study allows us 
to establish a number of features regarding
the non-equilibrium stationary state. The bulk thermal conductivity
and its dependence on the temperature over few decades was
found.
To our knowledge, such a computation from first principles,
without assuming linear response, has not been performed
previously. This was done using both linear response
(equilibrium) and direct (near equilibrium) approaches.
The $d=1$ Green-Kubo integrals in our theory are 
non-divergent and agree with the direct computation, in contrast
to  some of the other models.
We find that linear response theory is
quite robust and works well even when the steady state thermal
profile has significant curvature, and as a consequence we
derive an analytic description for the profile. By driving the
system farther from equilibrium, we are able to see when linear
response breaks down. Surprisingly, this is found to be 
near the same point where local equilibrium
becomes noticeably violated.
A sufficient condition for noticeable departures from linear
response and local equilibrium in $d=1,3$ seems to be $\lambda\nabla
  T/{T}\agt 1/10$.
It would be interesting to examine
the dynamics of non-equilibrium phase transitions or explore
dynamical cooling of boundary temperatures to provide a means to
access more complex non-equilibrium environments, and in
particular, the dynamics of ultrarelativistic heavy--ion
collisions.

We acknowledge support through 
grants at Keio University and DOE grant DE-FG02-91ER40608.

\begin{figure} 
\begin{center}
    \leavevmode
  \epsfysize=8cm\epsfbox{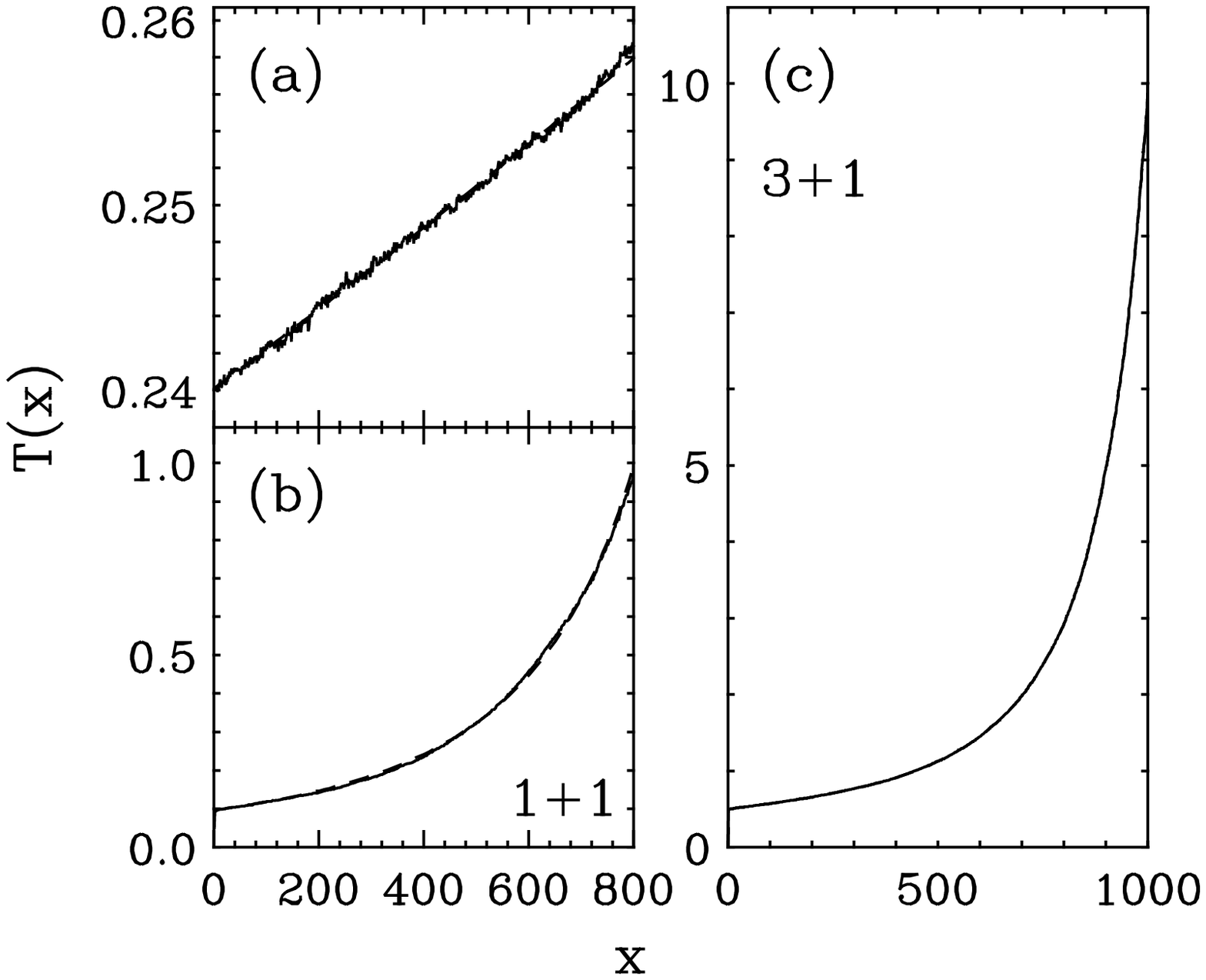}
  \caption{(a) Linear thermal profile (solid) near equilibrium
    for L=800, $d=1$ compared to a linear fit (dashes). The
    temperature was sampled at $\Delta t=1$ for $5\times 10^6$ points.
    (b) Sagging profile for a non-equilibrium steady state in (1+1)
    dimensions far from equilibrium (solid) compared to theoretical
    predictions (dashes), where $(T_1,T_2)=(0.1,1)$. The temperature
    was sampled $10^6$ times every $\Delta t=0.25$.  (c) A curved
    profile in (3+1) dimensions (solid) with the theoretical fit
    (dashes) for a $1000\times3\times3$ system with
    $(T_1,T_2)=(0.5,10)$.  The temperature was sampled $4\times10^6$
    times every $\Delta t=0.1$.  In both (b) and (c), the fits are
    virtually indistinguishable from the temperature profiles.  }
  \label{fig:one}
  \end{center}
\end{figure}

\begin{figure}
  \begin{center}
    \epsfxsize=8.6cm\epsfbox{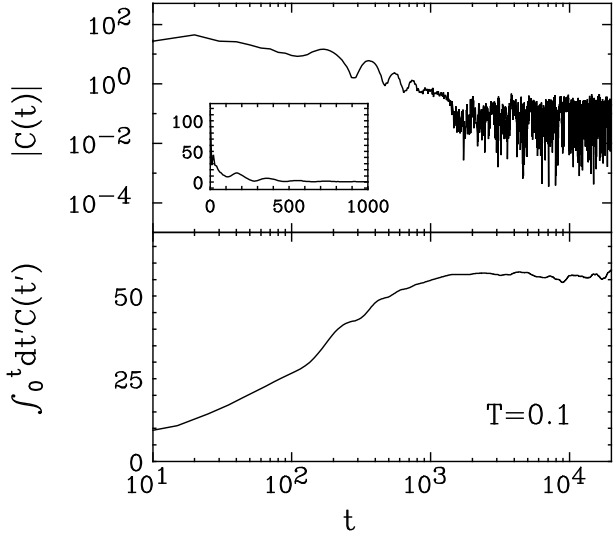}
    \caption{(Top) Time dependence of $|C(t)|$, where $C(t)=\int
      dx   \left\langle {\cal T}^{0x}(x,t)   {\cal
          T}^{0x}(x_0,0)\right\rangle_{\scriptscriptstyle EQ}/T^2$, for 
      $L=100$, up to a time $t=2\times 10^4$, well into the region
      where long-time tails should be evident. 
      Inset: Short time decorrelation. (Bottom) Green-Kubo integral up
      to time $t$, converging to $\kappa$ in 1+1 dimensions.}
    \label{fig:thr}
  \end{center}
\end{figure}

\begin{figure}
  \begin{center}
    \epsfxsize=8.5cm\epsfbox{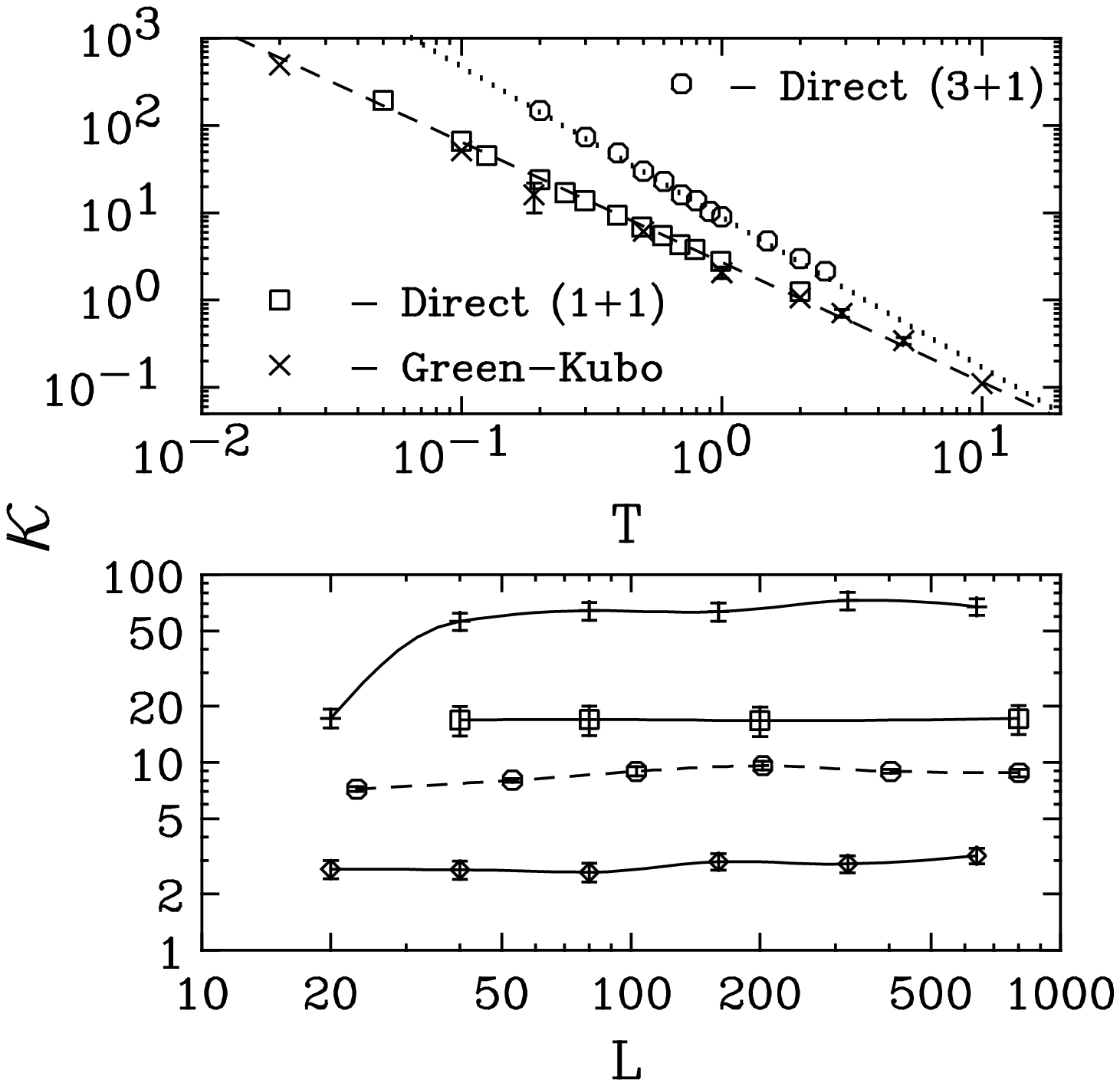}
    \caption{(Top) Thermal conductivity $\kappa$ obtained from
      direct ($\Box$ 1+1-d; $\circ$ 3+1-d) and Green-Kubo ($\times$)
      measurements for
      various lattice sizes $L$, and the power law fit
      (dashes,dots). (Bottom) $L$ dependence of $\kappa$ 
      indicating bulk behavior for temperatures (upper to lower)
      $T=1/10,1/4,1$ (solid, 1+1-d) and $T=1$ (dashes, 3+1-d).}
    \label{fig:two}
  \end{center}
\end{figure}
 
\begin{figure}
  \begin{center}
    \epsfysize=8cm\epsfbox{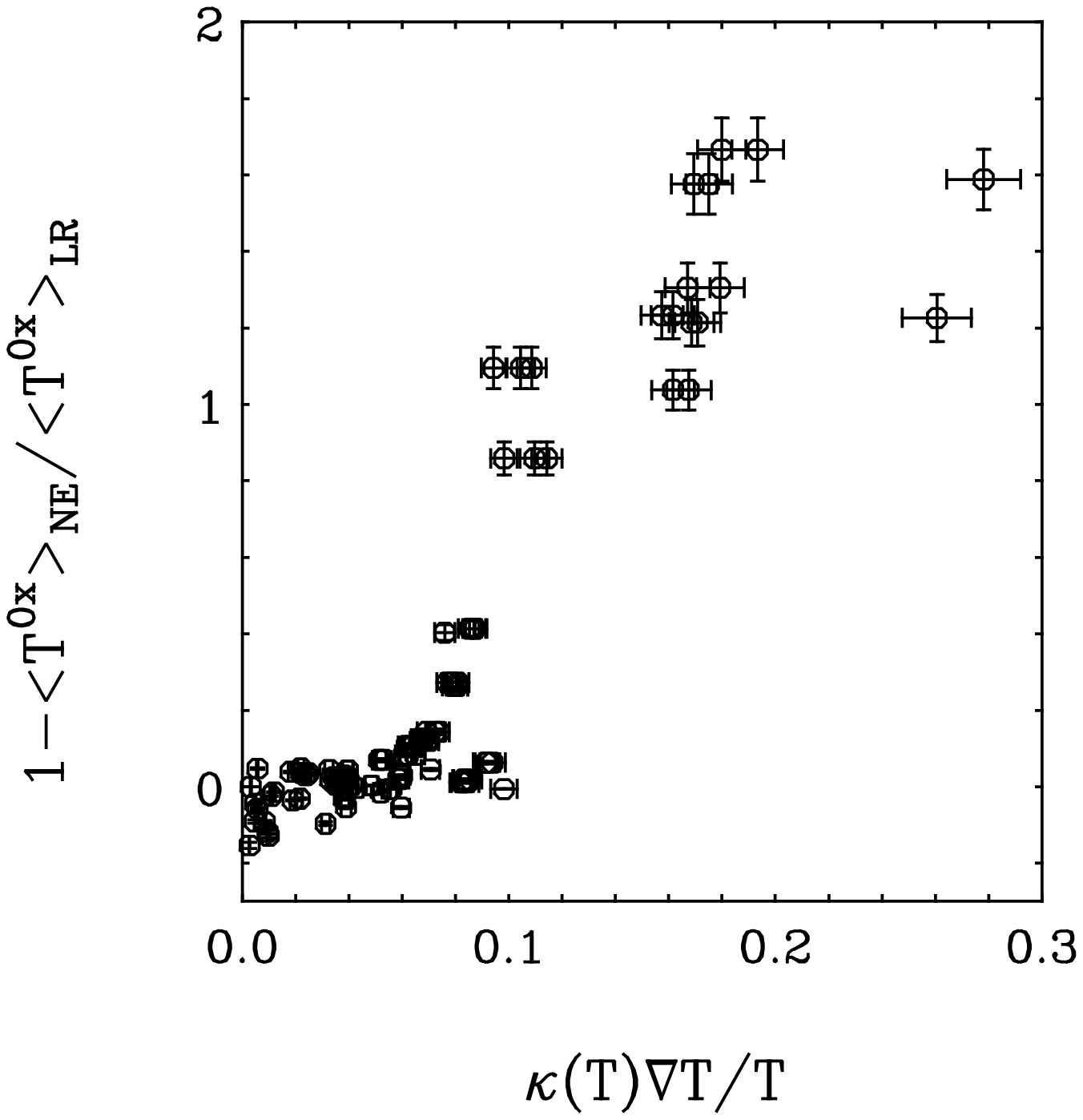}
    \caption{
      Breakdown of the linear response theory: The deviation of the
      measured energy flow $\langle T^{0x}\rangle_{NE}$ from
      its linear response expectation, $\langle
      T^{0x}\rangle_{LR}$, is shown as a function of $\kappa
      \nabla T/T$ for $d=1$ non-equilibrium steady
      states.  As discussed in the text, local equilibrium fails 
      simultaneously.
      }
    \label{fig:fou}
  \end{center}
\end{figure}

\begin{references}


\bibitem{revs} D.J.Evans, G.P.Morriss, {\sl Statistical
Mechanics of Non-Equilibrium Liquids} (Academic,~ London, 1990);
W.G.Hoover, {\sl Computational Statistical Mechanics} (Elsevier,
Amsterdam, 1991); P. Gaspard, {\sl Chaos, Scattering and 
Statistical Mechanics}, (Cambridge, New York, 1998).


\bibitem{gong} C. Gong, PhD Thesis, Duke University, (1994).

\bibitem{parisi} G. Parisi, {\sl  Europhys. Lett.} {\bf 40} (1997) 357.

\bibitem{ssb} L. Caiani, L. Casetti, M. Pettini, {\sl
  J. Phys. A: Math. Gen.} {\bf 31} (1998) 3357; L. Caiani {\it et
  al}, {\sl Phys. Rev. } {\bf E57} (1998) 3886.

\bibitem{mclennan} See for example, D. Zubarev, V. Morozov and 
G. R\"opke, {\it Statistical Mechanics of Non-equilibrium Processes}, 
(Akademie Verlag, Berlin, 1996); 
 J.A.~McLennan, {\sl Phys. Fluids} {\bf 4} (1961) 1319.

\bibitem{non-eq} W.G. Hoover, {\sl J. Chem. Phys.} {\bf 109} 4184 (1998).

\bibitem{sm}D.Yu.~Grigoriev, V.A.~Rubakov, {\sl Nucl. Phys. }{\bf B299} 
(1988) 67;
  K. Kajantie, M. Laine, K. Rummukainen, M. Shaposhnikov, {\sl
    Nucl. Phys.} {\bf
    B458} (1996) 90, {\bf B466} (1996) 189, and references therein.

\bibitem{classical-ym} 
  B. M\"uller, A. Trayanov,{\sl Phys. Rev. Lett. }{\bf 68} (1996);
  3387;
  C. Gong, {\sl Phys. Rev. }{\bf D49} (1994) 2642, and references therein.

\bibitem{hosoya} 
  Yu. S. Gangnus, A.V. Prozorkevich, S.A. Smolyanski\v i
  {\sl JETP Lett.} {\bf 28} (1978) 347;
  A. Hosoya, M. Sakagami, M. Takao, {\sl Ann. Phys. (NY) }
  {\bf 154} (1984) 229; 
  S. Jeon, \prd{\bf 52D} (1995) 3591; 
  S. Jeon, L. Yaffe, \prd{\bf 53D} (1996) 5799

\bibitem{smit}
  G. Aarts, J. Smit, {\sl Phys. Lett. }{\bf393B} (1997) 393; 
  {\sl Nucl. Phys.}{\bf B511} (1998) 451.

\bibitem{numrec} See, for instance, W.H. Press, B.P. Flannery,
  S.A. Teukolsky, W.T. Vetterling, {\sl Numerical Recipes},
  (Cambridge Univ. Press, New York, 1992).

\bibitem{thermostats} D. Kusnezov,  J. Sloan, {\sl
   Nucl. Phys.} {\bf B409} (1993) 635;  D. Kusnezov,  A. Bulgac, W. Bauer,
  {\sl Ann. Phys. }{\bf 204} (1990) 155.

\bibitem{tail} J.R. Dorfman, E.G.D. Cohen, {\sl
    Phys. Rev. Lett. }{\bf 25 } (1970) 1257;
  M.H.~Ernst,   E.H.~Hauge, J.M.J~van~Leeuwen, {\sl Phys. Rev.
    Lett. }{\bf 25} (1970) 1254, {\sl Phys. Rev. }{\bf A} 1971
    2055;
  Y.~ Pomeau, P. R\'esibois, {\sl Phys. Rep. }{\bf  19}
    (1975) 63.

\bibitem{others} S. Lepri, R. Livi, A. Politi, {\sl
Phys. Rev. Lett. }{\bf 78} (1997) 1896; H. Kaburaki, M. Machida,
{\sl Phys. Lett.} {\bf A181} (1993) 85; M. Mareschal, A. Amellal, {\sl
    Phys. Rev. }{\bf A37} (1988) 2189.

\bibitem{phonon}  C. Herring, {\sl Phys. Rev.} {\bf 95} (1954) 954.






\end{references}
\end{document}